\documentclass[conference]{IEEEtran}
\usepackage{hyperref}
\ifCLASSINFOpdf
\else
\fi
\usepackage{url}

\makeatletter
\renewcommand*{\thetable}{\arabic{table}}
\renewcommand*{\thefigure}{\arabic{figure}}
\let\c@table\c@figure
\makeatother 

\def\tablename{Table}

\usepackage{array}
\usepackage{color}
\usepackage{graphicx}
\usepackage{subfig}
\newcommand{\comment}[2]{}

\newcommand{\job}{j}
\newcommand{\Tasks}{T}
\newcommand{\prob}{p}
\newcommand{\freq}{f}
\newcommand{\importance}{s}

\newcommand{\onetonlinelong}{O*NET OnLine}

\newcommand{\onetonline}{O*NET}
\newcommand{\Onetonline}{O*NET}
\newcommand{\tsimilar}{related}
\newcommand{\jrelated}{related}

\newcommand{\ptask}[2]{t_{#1}^{#2}}
\newcommand{\tip}[2]{x_{#1}^{#2}}
\newcommand{\taskDiff}[4]{\ptask{#1, #3}{#2, #4}}
\newcommand{\tipDiff}[4]{\tip{#1, #3}{#2}}

\begin{document}
\title{Opening the Frey/Osborne Black Box: Which Tasks of a Job are Susceptible to Computerization?
}

\author{\IEEEauthorblockN{Philipp Brandes}
\IEEEauthorblockA{ETH Zurich, Switzerland\\
pbrandes@ethz.ch}
\and
\IEEEauthorblockN{Roger Wattenhofer}
\IEEEauthorblockA{ETH Zurich, Switzerland\\
wattenhofer@ethz.ch}
}


%Telephone: (800) 555--1212, Fax: (888) 555--1212}


\maketitle

\begin{abstract}
In their seminal paper, Frey and Osborne quantified the automation of jobs, by assigning each job in the \onetonline{} database a probability to be automated.
In this paper, we refine their results in the following way:
Every \onetonline{} job consists of a set of tasks, and  these tasks can be  \tsimilar.
We use a linear program to assign probabilities to tasks, such that \tsimilar{} tasks have a similar probability and the tasks can explain the computerization probability of a job.
Analyzing jobs on the level of tasks helps comprehending the results, as experts as well as laymen can more easily criticize and refine what parts of a job are susceptible to computerization.
% \philipp{A number from the eval chapter would be nice}
\end{abstract}


% \fi
\IEEEpeerreviewmaketitle

\section{Introduction}
% \philipp{our result vs our results, inconsistent -> change to results}

% Although newspapers and other popular media often claim the opposite, clock speeds of microprocessors have stopped increasing more than 10 years ago. 
Computerization is considered to be one of the biggest socio-economic challenges.
What is the foundation of the recent worries about many jobs being affected by automation~\cite{RaceAgainst,osbornefrey,SecondMachineAge,RiseOfRobots}? 

Why did the last few years see dramatic technological progress regarding self-driving cars~\cite{guizzo2011google}, board games~\cite{44806}, automatic language translation~\cite{auli2013joint}, or face recognition~\cite{le2013building}? 
One reason is big data. 
While ``intelligent algorithms'' in the past were restricted to learning from  data sets with a few thousand examples, we now have exabytes of data.
Learning becomes even more powerful if you combine big data with a highly parallel hardware, stirred by the success of graphics processing units (GPUs). 
However, both of these technological advancements needed to be harvested, and they are with the advent of so-called deep learning algorithms, which have blown the competition away, starting with voice recognition~\cite{deng2009deep}. 
As a consumer, you can already witness some of these advancements on your smartphone, but a lot more is to come soon. 
We believe that these advancements will revolutionize white collar work and (with a little help from sensors and robotics) also blue collar work. 
In contrast to previous waves of innovation, this time new emerging jobs might not be able to compensate jobs endangered by the new technology.

In their seminal paper, Frey and Osborne~\cite{osbornefrey} quantitatively study job automation, predicting that 47\% of US employment is at risk of automation.
In order to calculate this number, Frey and Osborne labeled 70 of the  702 jobs  from the \onetonlinelong{} job database\footnote{\Onetonline{} is an application that was created for the general public to provide broad access to the \onetonline{} database of occupational information. 
The site is maintained by the National Center for O*NET Development, on behalf of the U.S. Department of Labor, Employment and Training Administration (USDOL/ETA); 
see https://www.onetonline.org/} 
 manually as either ``automatable'' or  ``not automatable''.
Then, for the remainder of the jobs in the \onetonline{} database, they computed the automation probability as a function of the distance to the labeled jobs.

% We would like to go one step further. 
But the results of Frey and Osborne are opaque, one either believes their ``magic'' computerization percentages, or one has doubts.
We want anybody to be able to easily understand and argue about our results, by incorporating the unique tasks of each job.
This additional depth will help laymen as well as job experts to argue about potential flaws in our methodology.

% Every task of a job that is 100\% automatable must also be completely automatable.
If we know that a job is 100\% automatable, we also know that every task of that job must be completely automatable.
But what if a job is 87\% automatable?
Is every task 87\% automatable? Or are 87\% of the tasks completely automatable, and 13\% not at all?
% Is it a mix of these two extremes?
% Understanding this is one of the main goals of this paper:
We want to forecast which tasks of a job are safe and which tasks are automatable.

% We do not stop there.
In order to calculate the automation probability for a task, we first need to determine its share of a job (Section~\ref{sec:share}).
Based on this, we are able to assign each task a probability to be automated such that the weighted average of the probabilities is equal to the probability of the corresponding job (Section~\ref{sec:taskProbs}).
During our evaluation (Section~\ref{sec:eval}), we discover a few suspicious results in the probabilities by Frey and Osborne, e.g., a surprisingly high automation probability of 96\% for the job \emph{compensation and benefits managers}.
We conclude our paper by analyzing the correlation between various properties of a job and its probability to be automated (Section~\ref{sec:further}).
E.g., we show that there is a strong negative correlation between the level of education required for a job and its probability to be automated.
% \roger{maybe more?}

% We try to predict which tasks will be computerized next.
Our complete results can be found at  \url{http://jobs-study.ethz.ch}.

\section{Related Work}

The current effects of automation have been studied intensively in economics.
Most studies agree that some routine tasks have already fallen victim to automation~\cite{cookies,LousyJobs,ADPolarization}.
A task is routine if ``it can be accomplished by machines following explicit programmed rules''~\cite{cookies}.
% \cite{cookies}
With computers being able to do routine tasks, the demand for human labor performing these tasks has decreased.
But on the other hand, the demand for college educated labor has increased over the last decades~\cite{BJ92,W94,W98}.
The effect is more pronounced in industries that are computer-intensive~\cite{AKKComputers}.
As a consequence of this, the employment share of the highest skill quartile has increased.
In addition to more people being employed in the highest skill quartile, the real wage for this quartile has increased faster than the average real wage.
Service occupations, which are non-routine, but also not well paid, have also seen an increase in employment share and in real hourly wage.
Thus, both, employment share and real wage, are U-shaped with respect to the skill level~\cite{ADPolarization}.
This employment pattern is a phenomenon that is called polarization.
This is not unique to the US, but can also be observed, e.g., in the UK~\cite{LousyJobs}.
These papers make important observations about the effects that automation already has.
Until now, routine tasks are the ones most affected, but more and more tasks can nowadays be performed by a computer.
We focus on the future and try to predict which tasks will be automated next.

John Keynes predicted already in 1933 that there will be widespread technological unemployment ``due to the means of economising the use of labour outrunning the pace at which we can find new uses for labour''~\cite{Keynes1933}.
Automation might be the technology, where this becomes true~\cite{RaceAgainst,osbornefrey,SecondMachineAge,RiseOfRobots}.
``Automation of knowledge work'', ``Advanced robotics'', and ``Autonomous and near-autonomous vehicles'' are considered to be 3 out of 12 potentially economically disruptive technologies~\cite{MGI2}.
Computer labor and human labor may no longer be complements, but competitors.
Automation might be the cause for the current stagnation~\cite{RaceAgainst}.
There might be too much technological progress, which causes high unemployment.
A trend that could be going on for years, but was hidden by the housing boom~\cite{housingBoom}.

The seminal paper by Frey and Osborne is the first to make quantitative claims about the future of jobs~\cite{osbornefrey}.
Together with 70 machine learning experts, Frey and Osborne first manually labeled 70 out of 702 jobs from the \onetonline{} database as either ``automatable'' or ``non automatable''.
This labeling was, as the authors admit, a subjective assignment based on ``eye balling'' the job descriptions from \onetonline{}.
Labels were only assigned to jobs where the whole job was considered to be (non) automatable, and to jobs where the participants of the workshop were most confident.
% \philipp{Should we add: The authors do not explicitly the requirements for a job to be non-automatable.}
To calculate the probability for non-labeled jobs, Frey and Osborne used a probabilistic classification algorithm.
They chose 9 properties from  \onetonline{}  as features for their classifier, namely ``Finger Dexterity'', ``Manual Dexterity'', ``Cramped Work Space, Awkward Positions'', ``Originality'', ``Fine Arts'', ``Social Perceptiveness'', ``Negotiation'', ``Persuasion'', and ``Assisting and Caring for Others''.
% Repeatedly using cross validation with half the labeled data as test and half as training set yielded an average area under the curve of 0.894, which would be 1 for a perfect classifier and 0.5 for a random classifier.

% We believe that they are very useful and would like to point out that they were also the basis for the manual classification by Frey and Osborne.

The results from Frey and Osborne for the US job market were adopted to other countries, e.g., Finland, Norway, and Germany~\cite{jobsNorway,jobsGermany}.
This was done by matching each job from \onetonline{} to the locally used standardized name.
Due to differences in the economies, a different percentage of people will be affected by this change, e.g., only one third in Finland and Norway are at risk compared to 47\% in the US.
% We also use the probabilities of Frey and Osborne as a basis, but our goal is not to adopt it to one specific economy.
% We want to understand which tasks of a job are susceptible to computerization -- independent of the number of people performing a task.
% Instead of adopting the results to one country, we build upon them to deepen our understanding.

\section{Model}

\begin{figure}[htb]
 \centering
 \includegraphics[width=.75\linewidth]{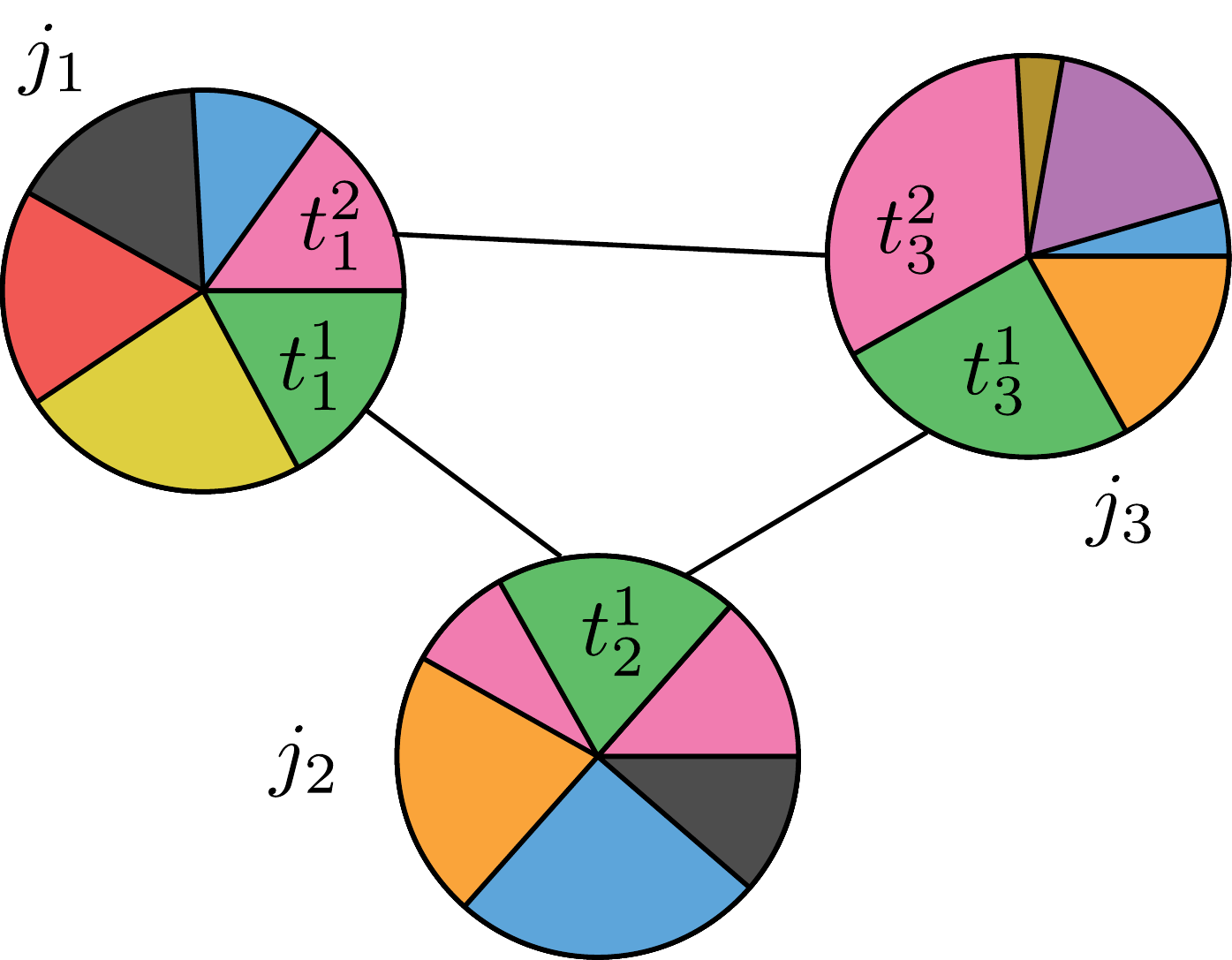}
 % pie-charts-neighbors.pdf: 546x600 pixel, 72dpi, 19.26x21.17 cm, bb=0 0 546 600
 \caption{This figure shows a small example consisting of three jobs represented by circles.
 The share of each task of a job is shown by its sector.
%  These indicate the share of each task. 
 Two tasks are connected by a line if and only if they are \tsimilar{}, i.e., similar according to \onetonline{}. Task $\ptask{1}{1}$ from job $\job_1$ and task $\ptask{2}{1}$ from job $\job_2$ are \tsimilar{} as indicated by the line connecting them. 
 Note that this relationship is not transitive. Thus, tasks $\ptask{1}{1}$ and $\ptask{3}{1}$ do not need to be \tsimilar{}.}
 \label{fig:model}
\end{figure}

We are given a set of jobs $J=\{\job_1,\ldots,\job_n\}$.
Each job $\job_i$ consists of a set of tasks $\Tasks_i=\{\ptask{i}{1}, \ldots, \ptask{i}{m}\}$, where every task belongs to exactly one job.
We call two tasks $\ptask{i}{k}$ and $\ptask{i'}{k'}$ \tsimilar{} if and only if these tasks are similar according to \onetonline.
Two jobs with \tsimilar{} tasks are also called \jrelated.
An example with 3 jobs is depicted in Figure~\ref{fig:model}.

% For every task $\ptask{i}{k}$ we have additional information.
\Onetonline{} provides us for each task $\ptask{i}{k}$  with the information how often it is performed.
This information was gathered by asking job incumbents and occupational experts.
The options are ``yearly or less'', ``more than yearly'', ``more than monthly'', ``more than weekly'', ``daily'', ``several times daily'', and ``hourly or more''.
\Onetonline{} provides a percentage for each of the 7 options.
We denote these frequencies of task $\ptask{i}{k}$ with $\freq_1(\ptask{i}{k}), \ldots, \freq_7(\ptask{i}{k})$.
Since these values are percentages, for every task $\ptask{i}{k}$ they sum up to 100\%, i.e., $\sum_{\ell = 1}^7{\freq_\ell(\ptask{i}{k})}=1$.

Each job $\job_i$ has a given probability $\prob(\job_i)$ to be automated.
We want to use $\prob(\job_i)$ to calculate a probability to be automated for each task of this job.
% .and to calculate the share of each task of this job such that the weighted average is equal to the probability of a job.

\section{From Task Frequencies to Task Shares}
\label{sec:share}

\begin{figure}[htb]
% \subfloat[51-4093.00]{
% 
% \includegraphics[width=.45\linewidth]{./51-419300PlatingandCoatingMachineSetters,Operators,andTenders,MetalandPlastic.pdf}
% 
% }
% \hfill
% \subfloat[51-6051.00]{
% 
% \includegraphics[width=.45\linewidth]{./51-605100Sewers,Hand.pdf}
% 
% }

\subfloat[Electro-Mechanical Technicians]{

\includegraphics[width=.45\linewidth]{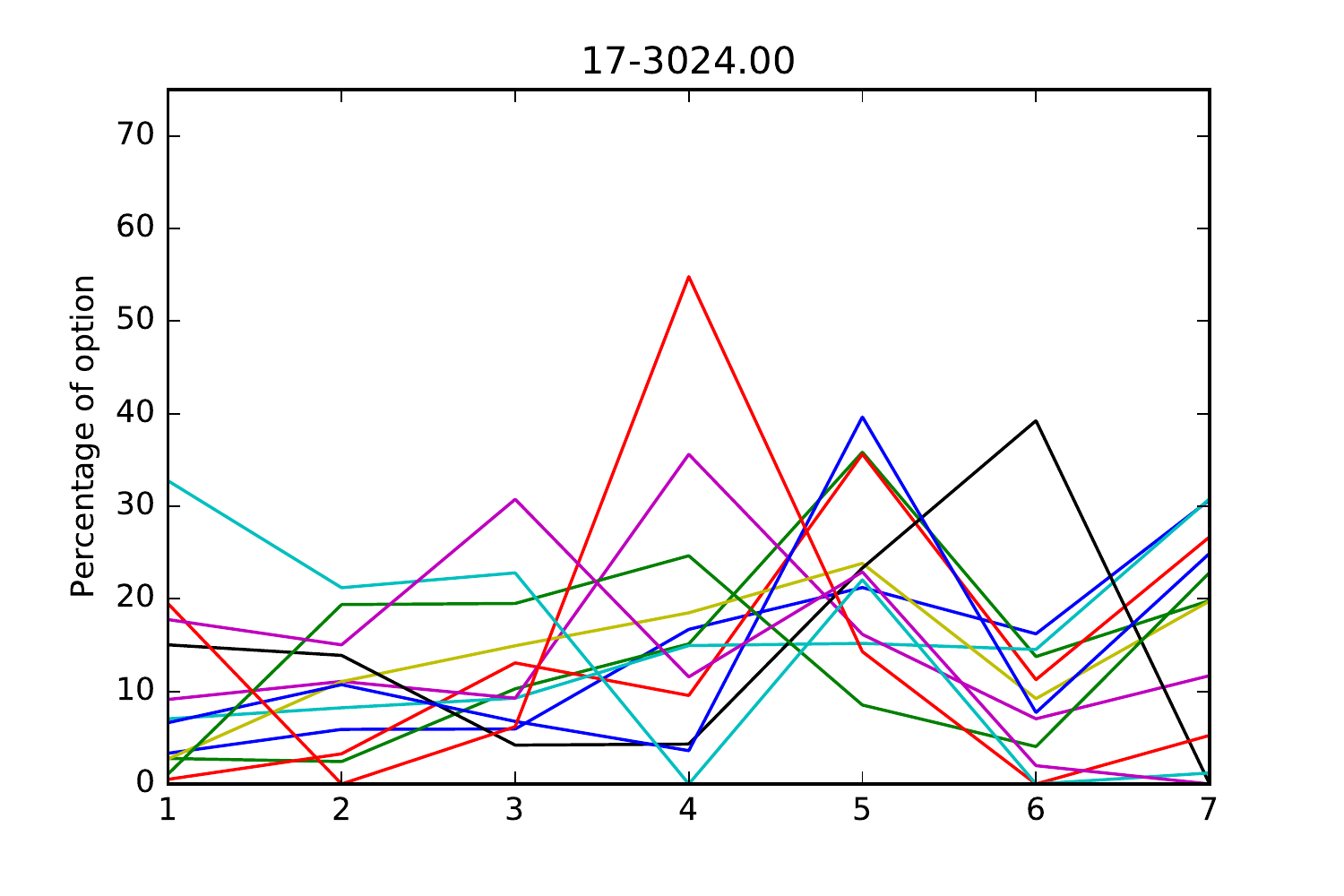}

}
\hfill
\subfloat[Environmental Engineering Technicians]{

\includegraphics[width=.45\linewidth]{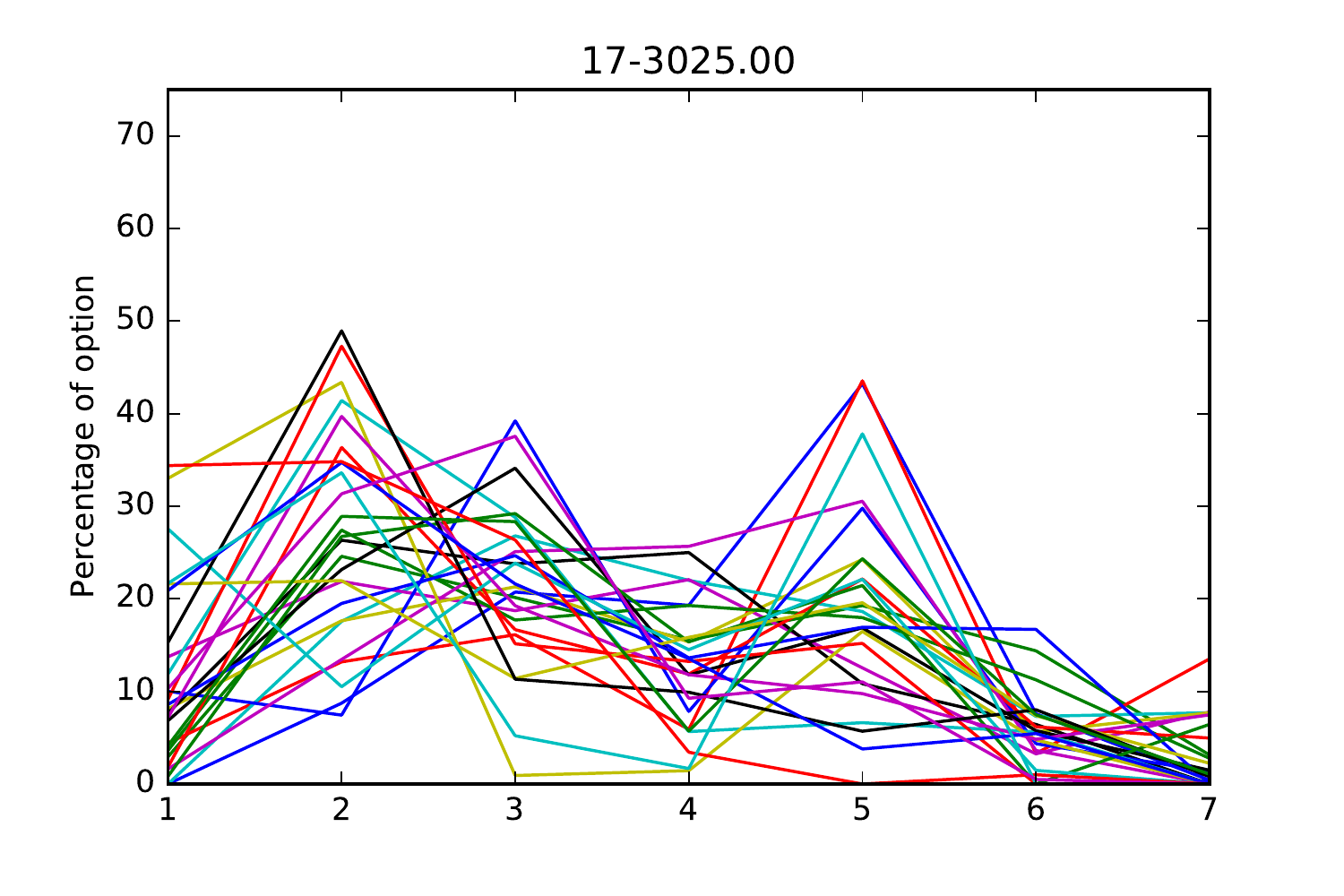}

}

\caption{The number of tasks and the frequencies assigned to them can differ significantly even for \jrelated{} jobs.
\label{fig:taskFrequencies}}

\end{figure}

We use the frequencies with which a task is performed to assign each task $\ptask{i}{k}$ its share $\importance(\ptask{i}{k})$.
For every task $\ptask{i}{k}$, the share denotes how much time is spent doing this task, such that $\sum_{\ptask{i}{k}\in\Tasks_i}\importance(\ptask{i}{k}) = 1$.
The frequency values from \onetonline{} do not fulfill this property; their values are very consistent for one job, but they can vary a lot between different jobs and might even seem to contradict each other.
An extreme example can be seen in Figure~\ref{fig:taskFrequencies}.
The seven frequency options provided by \onetonline{} are on the $x$-axis and on the $y$-axis is the corresponding value of each option.

To make use of the high consistency within a job,  we decided that the share of a task is a weighted average of its frequencies, i.e., $\importance(\ptask{i}{k}):=\sum_{\ell=1}^{7}\tip{i}{\ell} \freq_{\ell}(\ptask{i}{k})$.
We want to calculate the job specific coefficients $\tip{i}{\ell}$.
Let us illustrate these coefficients with a simple example.
If $\tip{i}{7} = 0.1$, then a task $\ptask{i}{k}$ that is done exclusively ``hourly or more'' (i.e., $\freq_7(\ptask{i}{k})=1$) makes up 10\% of job $\job_i$.

We want these coefficients to satisfy a few assumptions.
If \onetonline{} states that a task is done ``hourly or more'', then the share of this task should be higher than the share of a task that is done ``several times daily''.
This translates to $\tip{i}{\ell}\leq \tip{i}{\ell+1}$ $\forall \ell\in\{1,\ldots, 6\}$ and $0\leq \tip{i}{1}$.

These constraints neither use that jobs are \jrelated{} nor do they define the coefficients uniquely.
Both issues are solved if we require the coefficients $\tip{i}{\ell}$ and $\tip{i'}{\ell}$ for two \jrelated{} jobs $\job_{i}$ and $\job_{i'}$ to be similar.
The intuition behind this is that the frequencies of \onetonline{} for  \jrelated{} jobs are not independent of each other either, but rather should be similar as well.
Occupational experts who have rated the frequency in which a task is done for one job, are likely to have rated the frequencies of \jrelated{} jobs.

% I AM HERE

The coefficients cannot be identical without violating the other constraints.
Jobs have a different number of tasks and the frequencies are task specific.
The example in Figure~\ref{fig:taskFrequencies} highlights this.
It is therefore easy to see that we cannot have the same coefficients for two \jrelated{} jobs and fulfill the equality $\sum_{\ptask{i}{k}\in\Tasks_i}\importance(\ptask{i}{k}) = 1$ for both jobs simply because the number of tasks can differ a lot.

Thus, we allow a bit of slack in the coefficients of \jrelated{} jobs.
We use the variable $\tipDiff{i}{\ell}{i'}{\ell}$ to express the difference between the coefficients $\tip{i}{\ell}$ and $\tip{i'}{\ell}$ for two \jrelated{} jobs $\job_{i}, \job_{i'}$.
Formally, we define it as $\tipDiff{i}{\ell}{i'}{\ell} := \max\{\tip{i}{\ell}-\tip{i'}{\ell}, \tip{i'}{\ell}-\tip{i}{\ell} \}$.
This yields the following linear program, which minimizes the overall slack:
\newpage
\begin{eqnarray*}
\mbox{minimize} & \sum \tipDiff{i}{\ell}{i'}{\ell}\\
\mbox{s.t.}& & \\
\tipDiff{i}{\ell}{i'}{\ell} \geq& \tip{i}{\ell}  -\tip{i'}{\ell} &\forall \ell\\
& & \forall \job_i,\job_{i'}\in J  \mbox{ that are \jrelated{}}\\
	     \tipDiff{i}{\ell}{i'}{\ell}  \geq& \tip{i'}{\ell} -\tip{i}{\ell} &\forall \ell\\
& & \forall \job_i,\job_{i'}\in J  \mbox{ that are \jrelated{}}\\
 \sum_{\ptask{i}{k}\in T_i} \sum_{\ell=1}^{7}\tip{i}{\ell} \freq_{\ell}(\ptask{i}{k})  \leq& 1+\varepsilon &\forall \job_{i}\in J\\
  \sum_{\ptask{i}{k}\in T_i} \sum_{\ell=1}^{7}\tip{i}{\ell} \freq_{\ell}(\ptask{i}{k})  \geq& 1-\varepsilon &\forall \job_{i}\in J\\
\\
  \tip{i}{1}  \geq &0  &\forall \job_{i}\in J \\
  \tip{i}{\ell} \geq & \tip{i}{\ell-1} &\forall \job_{i}\in J \qquad \ell\in\{2,\ldots, 7\}
\end{eqnarray*}
We set $\varepsilon$ to 0.01.
% We would like to point out that we do not distinguish between jobs that have one pair of tasks that is \tsimilar{} and jobs that have several pairs.
% 
% 
The resulting LP has 169,372 variables in its objective function.
Since there are $735$ jobs,\footnote{We consider slightly more jobs than Frey and Osborne, since we use the finest granularity available from \onetonline{}.} this means that a job is \jrelated{} to approximately $32.9$ other jobs on average.
The value of the objective function is $24.6$, i.e., for two \jrelated{} jobs the coefficients differ only by $0.000145$ on average.
For comparison, the average value of a coefficient is $0.060$.
Our complete results can be found online at  \url{http://jobs-study.ethz.ch}.

\section{From Jobs to Tasks}
\label{sec:jobToTasks}

Knowing the shares of the  tasks enables us to set up a linear program that calculates for each task the probability  to be automated.
% We want the probabilities of the tasks to explain the probability of a job.
We want that the weighted average of the automation probabilities $p(\ptask{i}{k})$ of the tasks of a job $\job_i$ can explain the automation probability $\prob(\job_i)$ of the job, i.e., $\sum_{\ptask{i}{k}\in \Tasks_{i}}p\left(t_{i}^{k}\right)\cdot \importance\left(t_{i}^{k}\right)\approx p\left(j_{i}\right)$.
Furthermore, we want to assign \tsimilar{} tasks similar automation probabilities.
To do this, we define a variable $\taskDiff{i}{k}{i'}{k'}$ for each pair of \tsimilar{} tasks $\ptask{i}{k}$ and $\ptask{i'}{k'}$.
It denotes the probability difference  that we assign to the two tasks. 
Formally, it is defined as $\taskDiff{i}{k}{i'}{k'}:=\max\{ p(t_{i}^{k})-p(t_{i'}^{k'}),p(t_{i}^{k})-p(t_{i'}^{k'})\}$.
We want to minimize the sum of these variables, i.e., the sum of the probability difference of all \tsimilar{} tasks.

Combining these requirements with necessary conditions to have meaningful probabilities, i.e., $0\leq \prob(\ptask{i}{k})\leq 1$, yields the following linear program:

\begin{eqnarray*}
\mbox{minimize} & \sum \taskDiff{i}{k}{i'}{k'}\\
\mbox{s.t.} & &\\
  \prob(\ptask{i}{k})   -\prob(\ptask{i'}{k'})   \leq & \taskDiff{i}{k}{i'}{k'} & \forall  \ptask{i}{k}, \ptask{i'}{k'} \mbox{ that are \tsimilar}\\
  \prob(\ptask{i'}{k'})	-\prob(\ptask{i}{k}) 	 \leq & \taskDiff{i}{k}{i'}{k'} & \forall  \ptask{i}{k}, \ptask{i'}{k'} \mbox{ that are \tsimilar}\\
  \sum_{k}\prob(\ptask{i}{k})\cdot \importance\left(\ptask{i}{k}\right)  \leq& \prob(\job_{i})\left(1+\varepsilon\right) & \forall  \job_{i}\in J\\
  \sum_{k}\prob(\ptask{i}{k})\cdot \importance\left(\ptask{i}{k}\right)  \geq& \prob(\job_{i})\left(1-\varepsilon\right)& \forall  \job_{i}\in J\\
& &\\
  \prob(\ptask{i}{k})  \geq & 0 &\forall \job_{i}\in J, \ptask{i}{k}\in\Tasks_i\\ 
  \prob(\ptask{i}{k})  \leq & 1 &\forall \job_{i}\in J, \ptask{i}{k}\in\Tasks_i \\
  \taskDiff{i}{k}{i'}{k'}  \geq&0 & \forall  \taskDiff{i}{k}{i'}{k'}\\
  \taskDiff{i}{k}{i'}{k'}  \leq&1 & \forall  \taskDiff{i}{k}{i'}{k'}
\end{eqnarray*}
We set $\varepsilon$ to $0.01$.

\section{Linear Program Results}
\label{sec:results}
We now analyze the results of the linear program as described above.
Later on, we will look at a small refinement to automatically detect outliers in our results.

\subsection{Task Probabilities}
\label{sec:taskProbs}

\begin{figure}[htb]
 \centering
 \includegraphics[width=0.9\linewidth]{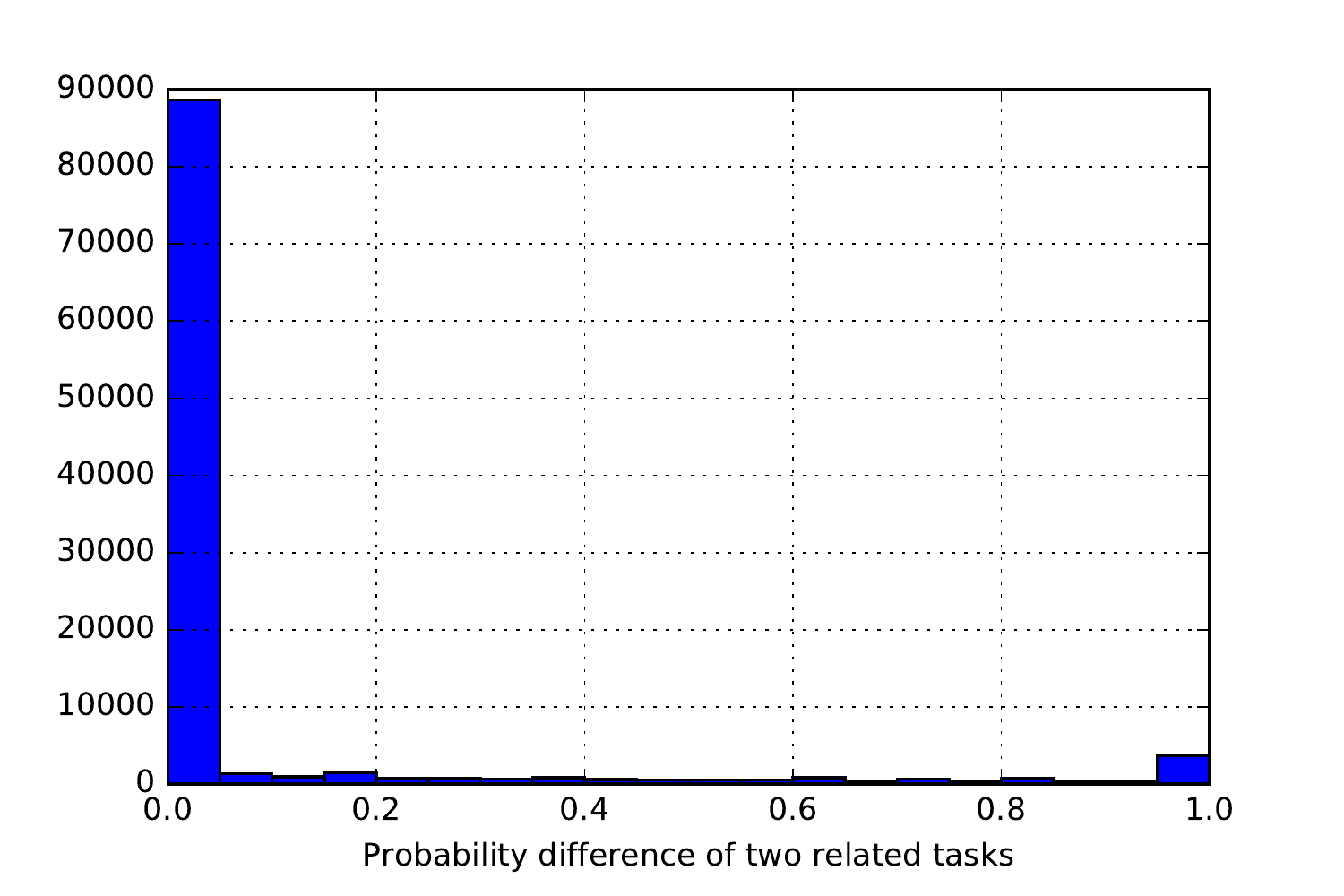}
 % probDiffTask.pdf: 432x288 pixel, 72dpi, 15.24x10.16 cm, bb=0 0 432 288
 \caption{A histogram of the probability with which two \tsimilar{} tasks differ.
 }
 \label{fig:taskDiffDistribution}
\end{figure}

The linear program as described in Section~\ref{sec:jobToTasks} has 105,748 variables in its objective function and it has a minimal value of 9,846.
This means that two \tsimilar{} tasks differ, on average, with regard to their probability by $9.3\%$.
The complete results can be found online at \url{http://jobs-study.ethz.ch}.
The histogram of the probability difference between two \tsimilar{} tasks is shown in Figure~\ref{fig:taskDiffDistribution}.
A majority of \tsimilar{} tasks  is assigned a similar probability.
A  small fraction of \tsimilar{} tasks is assigned diametrically opposed probabilities, which seems startling.
It can be reconciled by considering that neither the classification by Frey and Osborne nor the classification of tasks being \tsimilar{}  by \onetonline{} are perfect.

One example that highlights this are the two jobs \emph{computer programmer} and \emph{software developers, applications}.
These two jobs have many \tsimilar{} tasks, but the probabilities of these jobs  differ a lot (4\% for \emph{software developers, applications}, 48\% for \emph{computer programmers}).
Hence, the diametrically opposed probabilities are necessary to meet the constraints of the linear program.

In the following, we present a few selected jobs to illustrate our results.
The first example is \emph{chemists}.
This job has an automation probability of 10\% according to Frey and Osborne.
% Each task with its automation probability and its share are shown in Table~\ref{tab:chemist}.
Only one task has, according to our linear program,  a high probability of being automated: ``Induce changes in composition of substances by introducing heat, light,
energy, or chemical catalysts for quantitative or qualitative analysis.''
Other simple mechanical tasks have been assigned low automation probabilities.
We will revisit this job in Section~\ref{sec:eval}.

Next up: \emph{judges}.
Their automation probability is 40\%.
The tasks, their probabilities, and their shares are shown in Table~\ref{tab:judges}.
The tasks that can be automated can be grouped in two sets: 
preliminary hearings which includes making first assessments, and ensuring that the procedures in court are followed.
The tasks that involve sentencing (or the preparation thereof) have been assigned low automation probabilities.

\begin{table}
\scriptsize
\begin{tabular}{|>{\centering}p{6.5cm}|>{\centering}p{0.5cm}|>{\centering}p{0.5cm}|}
\hline 
Task Description & $\prob$ & Share\tabularnewline
\hline 
\hline 
Write decisions on cases.  & 1 & 5.1\tabularnewline
\hline 
Instruct juries on applicable laws, direct juries to deduce the facts
from the evidence presented, and hear their verdicts.  & 1 & 3.4\tabularnewline
\hline 
Monitor proceedings to ensure that all applicable rules and procedures
are followed.  & 1 & 8.0\tabularnewline
\hline 
Advise attorneys, juries, litigants, and court personnel regarding
conduct, issues, and proceedings.  & 1 & 6.2\tabularnewline
\hline 
Interpret and enforce rules of procedure or establish new rules in
situations where there are no procedures already established by law.  & 1 & 5.4\tabularnewline
\hline 
Conduct preliminary hearings to decide issues such as whether there
is reasonable and probable cause to hold defendants in felony cases.  & 1 & 3.9\tabularnewline
\hline 
Rule on admissibility of evidence and methods of conducting testimony.  & 0.94 & 5.3\tabularnewline
\hline 
Preside over hearings and listen to allegations made by plaintiffs
to determine whether the evidence supports the charges.  & 0.46 & 5.9\tabularnewline
\hline 
Perform wedding ceremonies.  & 0.39 & 2.7\tabularnewline
\hline 
Read documents on pleadings and motions to ascertain facts and issues.  & 0 & 10.1\tabularnewline
\hline 
Research legal issues and write opinions on the issues.  & 0 & 6.5\tabularnewline
\hline 
Settle disputes between opposing attorneys.  & 0 & 4.6\tabularnewline
\hline 
Participate in judicial tribunals to help resolve disputes.  & 0 & 6.6\tabularnewline
\hline 
Rule on custody and access disputes, and enforce court orders regarding
custody and support of children.  & 0 & 6.3\tabularnewline
\hline 
Sentence defendants in criminal cases, on conviction by jury, according
to applicable government statutes.  & 0 & 4.0\tabularnewline
\hline 
Grant divorces and divide assets between spouses.  & 0 & 4.7\tabularnewline
\hline 
Award compensation for damages to litigants in civil cases in relation
to findings by juries or by the court.  & 0 & 3.8\tabularnewline
\hline 
Supervise other judges, court officers, and the court's administrative
staff.  & 0 & 8.5\tabularnewline
\hline 
\end{tabular}

\caption{The automation probability and the share of each task of \emph{Judges, Magistrate Judges, and Magistrates"}. The automation probability of this job is 40\%.}
\label{tab:judges}

\end{table}

\begin{figure}[htb]
 \centering
 \includegraphics[width=.9\linewidth]{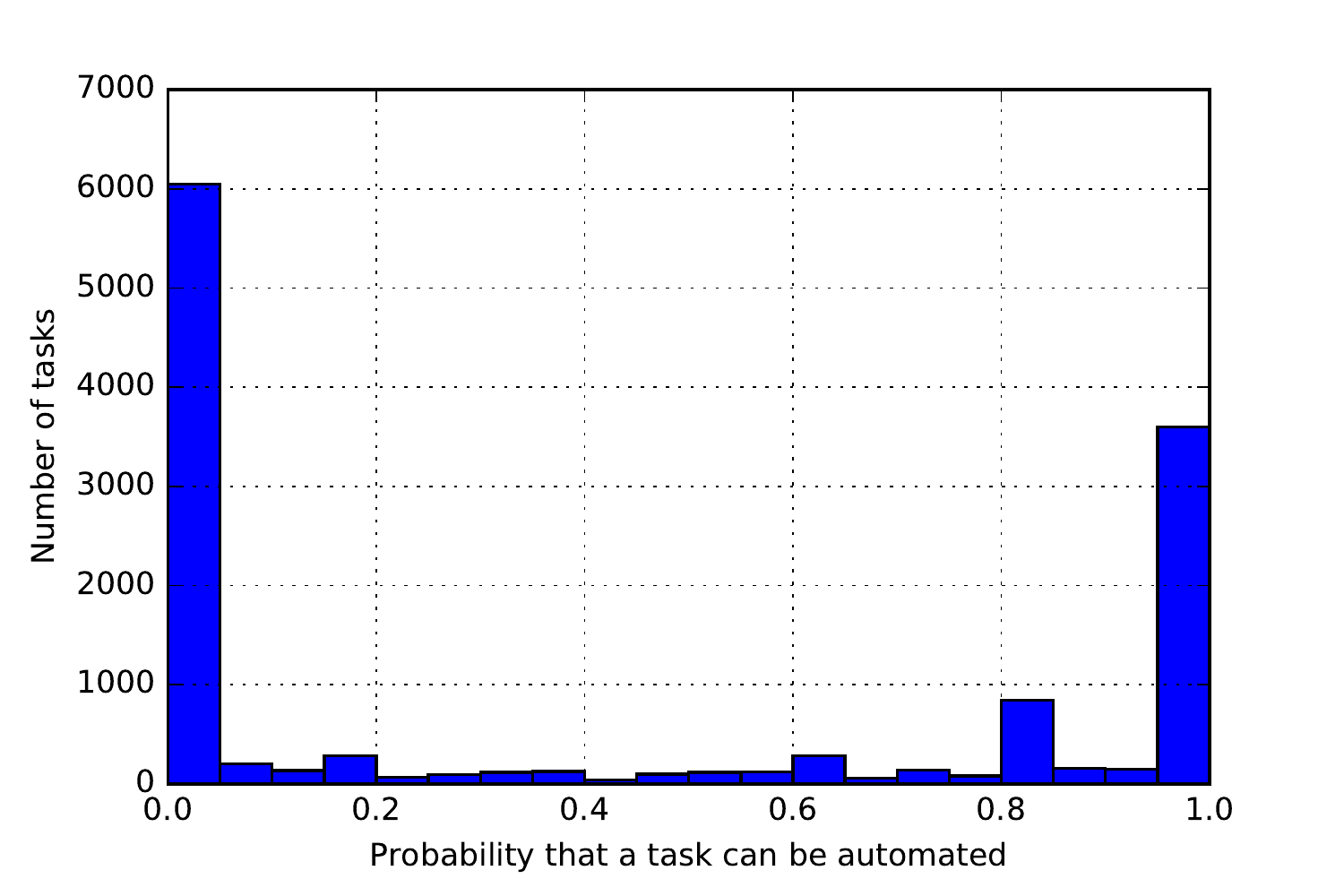}
 % task_prob_distribution.pdf: 432x288 pixel, 72dpi, 15.24x10.16 cm, bb=0 0 432 288
 \caption{A histogram of the probabilities that tasks can be automated. Nearly all tasks are assigned either 0 or 1. 
 This simplifies arguing whether our classification is correct.}\label{fig:taskdistribution}
\end{figure}

Figure~\ref{fig:taskdistribution} shows the histogram of the probabilities from our linear program.
The probabilities for most tasks are either very high or very low and only a few tasks have a probability in-between.
% This phenomenon can be observed across all jobs.
This desired side effect of our linear program helps us to achieve our goal of allowing job experts (and laymen) to argue about the validity of our results.
We invite the reader to have a look at other jobs at \url{http://jobs-study.ethz.ch}.

\subsection{Outlier Detection}
\label{sec:eval}
To evaluate our approach and check for outliers, we use a variant of cross-validation.
For every job $\job_i$, we create a linear program without job $\job_i$.
This yields a probability $\prob_i(\cdot)$ for every task but the tasks from $\job_i$.
Afterward, we calculate the new probability $\prob'(\job_i)$ that job $\job_i$ can be automated.
We do this by setting the probability $\prob'(\ptask{i}{k})$ of each task $\ptask{i}{k}$ to the average of all tasks that are \tsimilar{} to it.
We denote the set of \tsimilar{} tasks by $N(\ptask{i}{k})$.
Formally, we set $\prob'(\ptask{i}{k}):=\frac{1}{|N(\ptask{i}{k})|}\sum_{\ptask{i'}{k'}\in N(\ptask{i}{k})}\prob_i(\ptask{i'}{k'})$.

\begin{figure}[htb]
 \centering
 \includegraphics[width=1\linewidth]{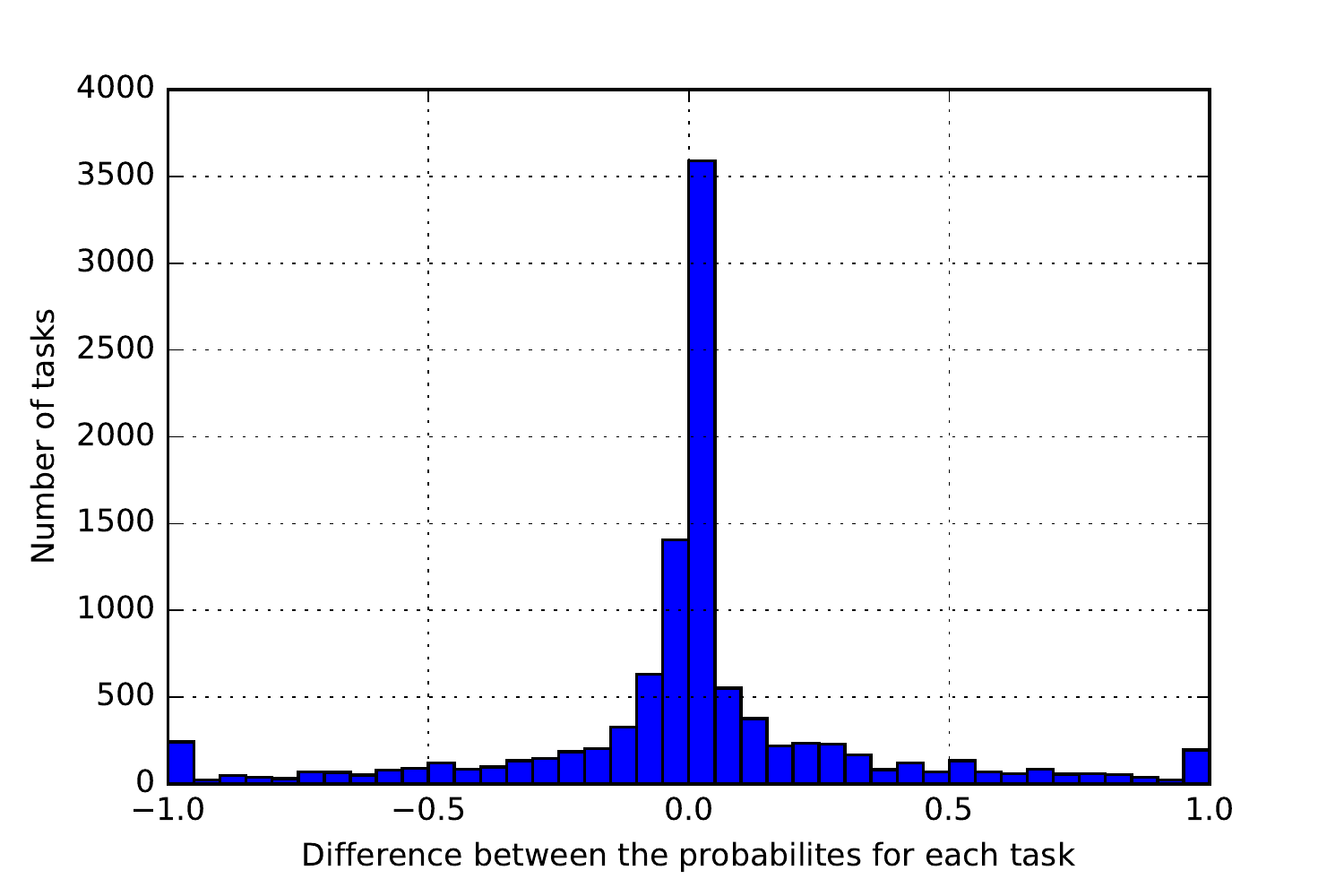}
 % taskComparative.pdf: 432x288 pixel, 72dpi, 15.24x10.16 cm, bb=0 0 432 288
 \caption{A histogram of $\prob(\ptask{i}{k})-\prob'(\ptask{i}{k})$ for all tasks. 
 This plot shows that most tasks have a similar probability in both approaches. 
 The values are sharply distributed around 0.}
 \label{fig:taskComp}
\end{figure}

We first compare $\prob'(\ptask{i}{k})$ with $\prob(\ptask{i}{k})$.
The difference between these two probabilities should be small for the majority of the tasks.
This is indeed what can be seen in Figure~\ref{fig:taskComp}.
The histogram of $\prob'(\ptask{i}{k}) - \prob(\ptask{i}{k})$ shows that nearly all tasks have similar probabilities in both approaches.
The average absolute difference is less than $20\%$.
The distribution is centered around 0. 
Its mean is less than $0.05\%$.

By combining the new probability of each task with its share, we can calculate the new probability of job $\job_i$ by using a weighted average.
This allows us to compare $\prob'(\job_i)$ with $\prob(\job_i)$.

\begin{figure}[htb]
 \centering
 \includegraphics[width=1\linewidth]{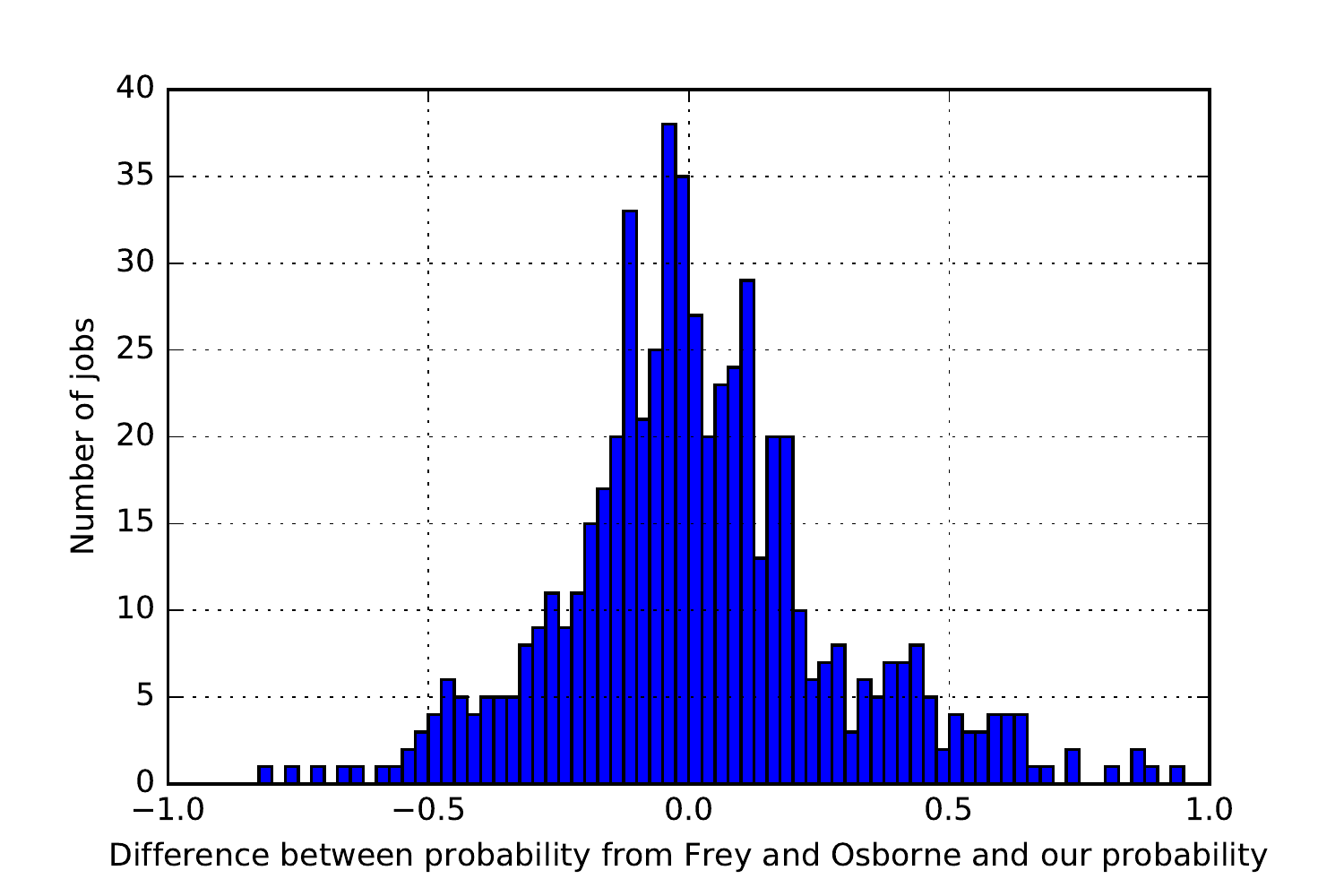}
 % probHistDiff.pdf: 432x288 pixel, 72dpi, 15.24x10.16 cm, bb=0 0 432 288
 \caption{A histogram of the difference between the probability by Frey and Osborne with our probability. 
 The distribution is centered around 0 and a majority of the jobs differs by less than 20\%.}
 \label{fig:probHist}
\end{figure}

We have plotted this difference, i.e., $\prob(\job_i)-\prob'(\job_i)$, in Figure~\ref{fig:probHist}.
We can see that the difference is centered around 0\%; with the average  absolute difference being less than $20\%$.
For more than half of the jobs our probability differs by less than 20\% from Frey and Osborne~\cite{osbornefrey}.
Most interesting are the jobs whose probability differs significantly.
We now have a look at a few of them.

There are jobs where our probability is more than 80\% smaller than the one by Frey and Osborne.
One job  is \emph{compensation and benefits managers}.
We assigned it a probability $\prob'(\job)$ to be automated of $9.1\%$; compared to $\prob(\job)=96\%$ by Frey and Osborne.
We do not claim to know the true value, but we can look at the job and compare it to the probabilities of jobs we consider similar.
Notice that this is conceptually similar to what our linear program does and thus might be biased.
The tasks of this job are shown in Table~\ref{tab:compMan}.
If we manually compare them to \tsimilar{} tasks,  we conclude that they do not seem to be automatable in the next few decades.
We do favor our result over the result of Frey and Osborne.

\begin{table}

\scriptsize

\begin{tabular}{|>{\centering}p{6.5cm}|c|c|}
\hline 
Task Description & $\prob$ & $\prob'$\tabularnewline
\hline 
\hline 
Advise management on such matters as equal employment opportunity,
sexual harassment and discrimination.  & 1 & 0.15\tabularnewline
\hline 
Study legislation, arbitration decisions, and collective bargaining
contracts to assess industry trends.  & 1 & 0\tabularnewline
\hline 
Fulfill all reporting requirements of all relevant government rules
and regulations, including the Employee Retirement Income Security
Act (ERISA).  & 1 & 0.20\tabularnewline
\hline 
Investigate and report on industrial accidents for insurance carriers.  & 1 & 0.12\tabularnewline
\hline 
Represent organization at personnel-related hearings and investigations.  & 1 & 0\tabularnewline
\hline 
Analyze compensation policies, government regulations, and prevailing
wage rates to develop competitive compensation plan.  & 1 & 0.5\tabularnewline
\hline 
Mediate between benefits providers and employees, such as by assisting
in handling employees' benefits-related questions or taking suggestions.  & 1 & 0.42\tabularnewline
\hline 
Prepare detailed job descriptions and classification systems and define
job levels and families, in partnership with other managers.  & 1 & 0\tabularnewline
\hline 
Prepare personnel forecasts to project employment needs.  & 1 & 0\tabularnewline
\hline 
Direct preparation and distribution of written and verbal information
to inform employees of benefits, compensation, and personnel policies.  & 1 & 0\tabularnewline
\hline 
Manage the design and development of tools to assist employees in
benefits selection, and to guide managers through compensation decisions.  & 1 & 0\tabularnewline
\hline 
Design, evaluate and modify benefits policies to ensure that programs
are current, competitive and in compliance with legal requirements.  & 1 & 0\tabularnewline
\hline 
Administer, direct, and review employee benefit programs, including
the integration of benefit programs following mergers and acquisitions.  & 1 & 0\tabularnewline
\hline 
Prepare budgets for personnel operations.  & 1 & 0.03\tabularnewline
\hline 
Maintain records and compile statistical reports concerning personnel-related
data such as hires, transfers, performance appraisals, and absenteeism
rates.  & 1 & 0\tabularnewline
\hline 
Contract with vendors to provide employee services, such as food services,
transportation, or relocation service.  & 1 & 0.38\tabularnewline
\hline 
Identify and implement benefits to increase the quality of life for
employees, by working with brokers and researching benefits issues.  & 1 & 0\tabularnewline
\hline 
Plan, direct, supervise, and coordinate work activities of subordinates
and staff relating to employment, compensation, labor relations, and
employee relations.  & 1 & 0\tabularnewline
\hline 
Negotiate bargaining agreements.  & 1 & 0.67\tabularnewline
\hline 
Plan and conduct new employee orientations to foster positive attitude
toward organizational objectives.  & 1 & 0\tabularnewline
\hline 
Conduct exit interviews to identify reasons for employee termination.  & 1 & 0\tabularnewline
\hline 
Develop methods to improve employment policies, processes, and practices,
and recommend changes to management.  & 0.51 & 0\tabularnewline
\hline 
Formulate policies, procedures and programs for recruitment, testing,
placement, classification, orientation, benefits and compensation,
and labor and industrial relations.  & 0.23 & 0.01\tabularnewline
\hline 
\end{tabular}
\caption{The tasks and their corresponding probability that they will be automated
for \emph{Compensation and Benefits Managers} according to our original linear program and the cross-validation.
\label{tab:compMan}}
\end{table}

There is only one job that we assign a much higher probability than Frey and Osborne.
The job \emph{First-Line supervisors of production and operating workers} has been assigned a 83\% automation probability by us and only 1.6\% by Frey and Osborne.
A close inspection of the tasks makes us believe that the true value is between these extremes.
Quite a few of the tasks are clearly automatable, e.g., ``Keep records of employees' attendance and hours worked.'' and ``Observe work and monitor gauges, dials, and other indicators to ensure that operators conform to production or processing standards.''
Others, e.g., ``Read and analyze charts, work orders, production schedules, and other records and reports to determine production requirements and to evaluate current production estimates and outputs.'' seem difficult to automate.
% We do not know the true probability, but we are inclined to think that it is closer to 80\% than to 0\%.
The complete results for this can job be found at \url{http://jobs-study.ethz.ch}.

We continue by comparing the previous results with the approach described in this section.
To do this, we return to the jobs that we have looked at previously.
First off is the job \emph{chemists}.
As shown in Table~\ref{tab:chemist2}, the automation probability of most tasks has increased.
Consequently, the automation probability of this job has increased from 10\% to 42\%.
Due to the large difference, this job should be analyzed in-depth by job experts.

\begin{table}
\scriptsize
\begin{tabular}{|>{\centering}p{7cm}|>{\centering}p{0.5cm}|>{\centering}p{0.5cm}|}
\hline 
Task Description & $\prob$ & $\prob'$\tabularnewline
\hline 
\hline 
Induce changes in composition of substances by introducing heat, light,
energy, or chemical catalysts for quantitative or qualitative analysis.  & 0.68 & 0.82\tabularnewline
\hline 
Analyze organic or inorganic compounds to determine chemical or physical
properties, composition, structure, relationships, or reactions, using
chromatography, spectroscopy, or spectrophotometry techniques.  & 0.18 & 0.82\tabularnewline
\hline 
Maintain laboratory instruments to ensure proper working order and
troubleshoot malfunctions when needed.  & 0.16 & 0.78\tabularnewline
\hline 
Conduct quality control tests.  & 0.07 & 0.54\tabularnewline
\hline 
Write technical papers or reports or prepare standards and specifications
for processes, facilities, products, or tests.  & 0.03 & 0.03\tabularnewline
\hline 
Study effects of various methods of processing, preserving, or packaging
on composition or properties of foods.  & 0 & 0.20\tabularnewline
\hline 
Prepare test solutions, compounds, or reagents for laboratory personnel
to conduct tests.  & 0 & 0.63\tabularnewline
\hline 
Purchase laboratory supplies, such as chemicals, when supplies are
low or near their expiration date.  & 0 & 1\tabularnewline
\hline 
Evaluate laboratory safety procedures to ensure compliance with standards
or to make improvements as needed.  & 0 & 0\tabularnewline
\hline 
Direct, coordinate, or advise personnel in test procedures for analyzing
components or physical properties of materials.  & 0 & 0.01\tabularnewline
\hline 
Develop, improve, or customize products, equipment, formulas, processes,
or analytical methods.  & 0 & 0\tabularnewline
\hline 
Confer with scientists or engineers to conduct analyses of research
projects, interpret test results, or develop nonstandard tests.  & 0 & 0.02\tabularnewline
\hline 
\end{tabular}
\caption{The automation probability and the share of each task of a \emph{chemist}.}
\label{tab:chemist2}
\end{table}

% talk about groups of tasks
The changes in the automation probability of the tasks of \emph{judges} are much smaller.
Most tasks have a similar automation probability as before and the overall probability of this job has changed marginally, i.e., increased only from 40\% to 50\%.
Therefore, we are confident that the classification by Frey and Osborne is correct.

We conclude that our approach can also be used to 
% Setting the probability of a task to the average of its neighbors allows us to
detect outliers in the results of Frey and Osborne.
We can then manually inspect the automation probabilities of the tasks of such an outlier to determine the truth.
We think our results allow us to fine tune the results from Frey and Osborne, but not replace it, as we need their results to bootstrap our linear program.
% Without their initial, very good classification, our results would not be possible.
% We would like to point out that the approach described in this section cannot be applied iteratively, i.e., using the new probabilities as a basis and reiterate the whole process.
% Since we assign each tasks the average of its neighbors, these probabilities would converge to a value around 50\% after sufficiently many iterations.

\section{Further Analysis}
\label{sec:further}

\begin{figure*}[ht]
\subfloat[Deductive Reasoning]{

\includegraphics[width=.45\linewidth]{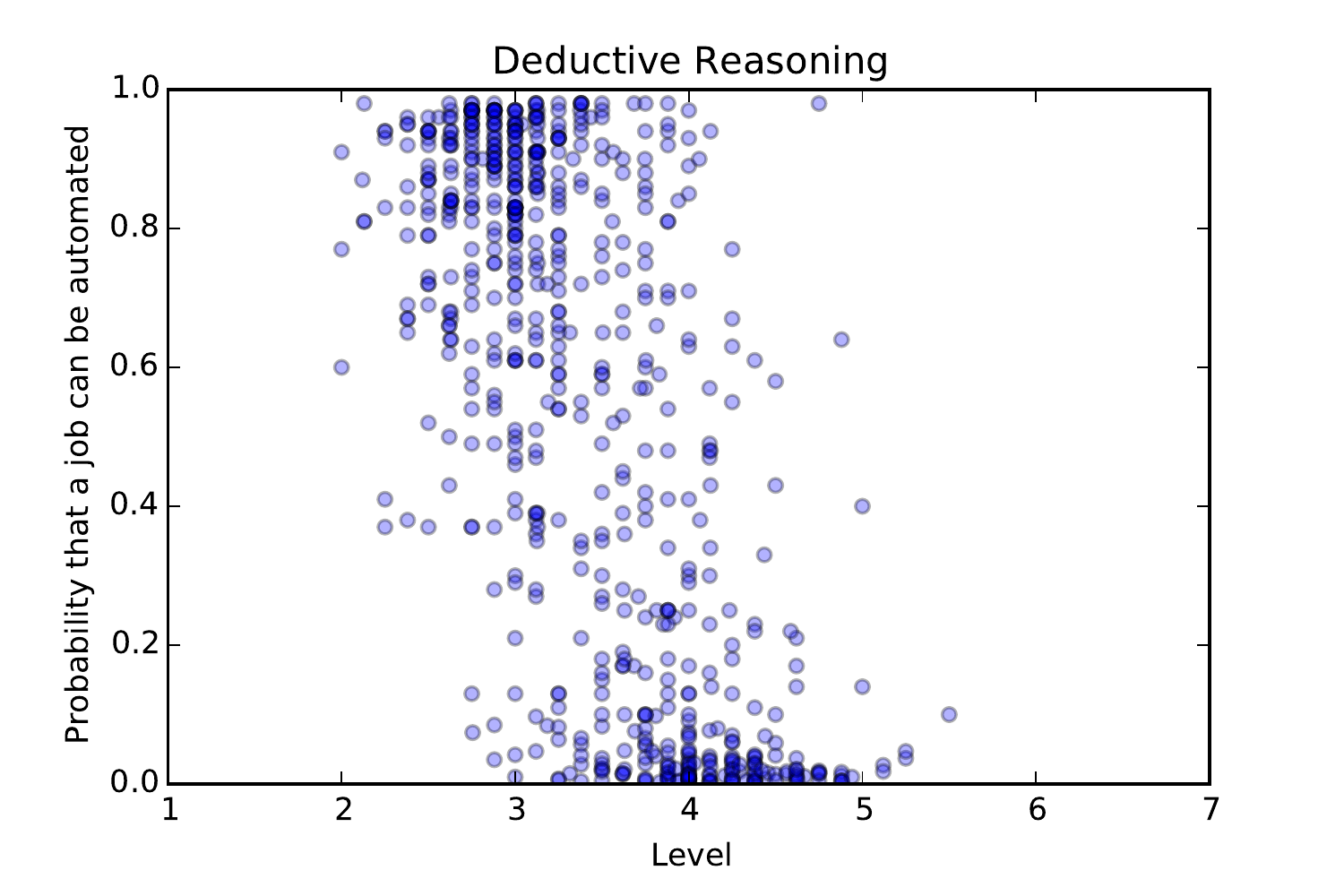}

}
\hfill
\subfloat[Originality]{

\includegraphics[width=.45\linewidth]{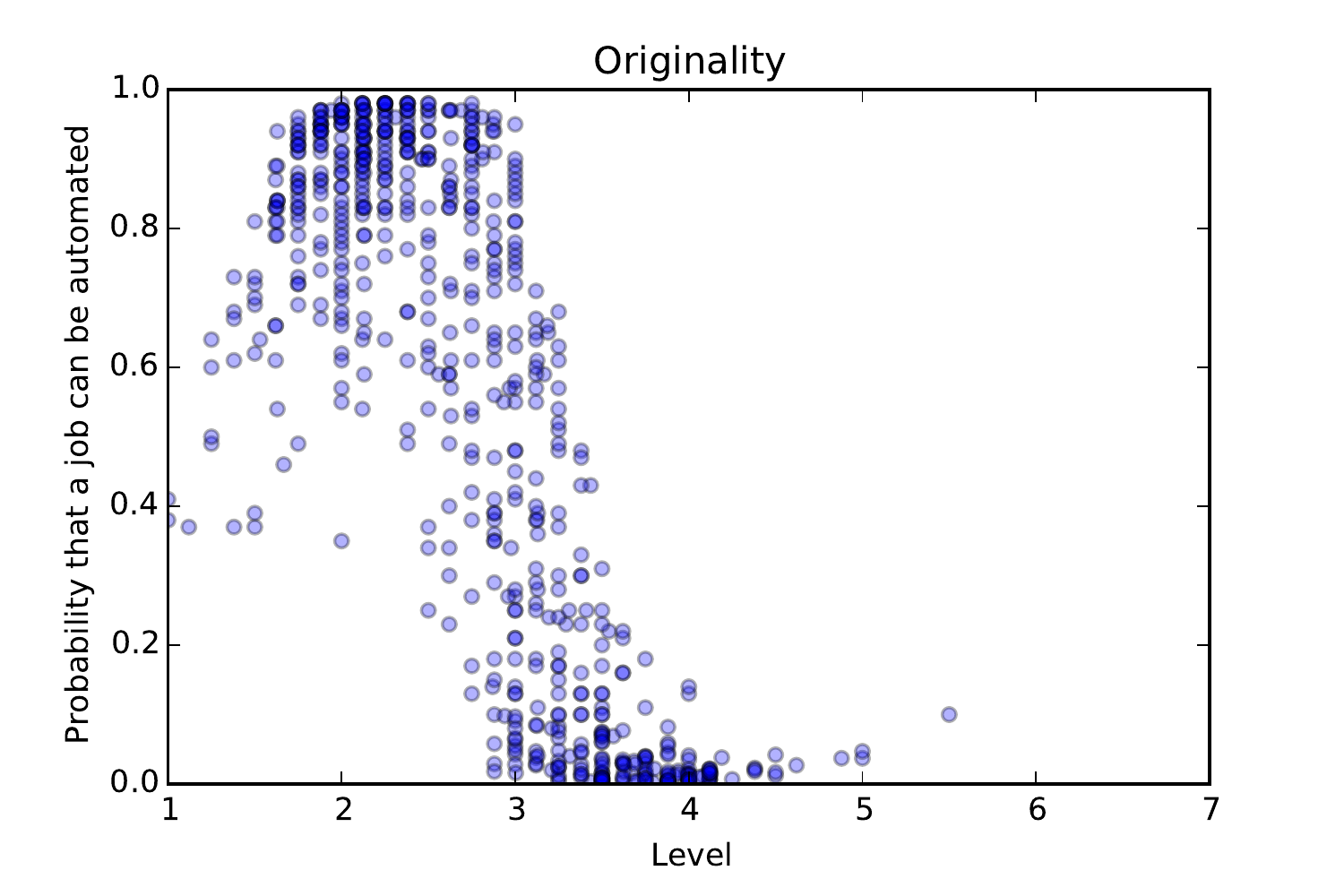}

}
% %
% \hfill
% \subfloat[Thinking Creatively]{
% %
% \includegraphics[width=.3\linewidth]{./LVThinkingCreatively.pdf}
% 
% }

\caption{The probability that a job can be automated over the level of the  abilities ``deductive reasoning'' and  ``originality'' used in this job.
These levels are defined by \onetonline{} and for each they provide an anchor point.
Level 2 of ``deductive reasoning'' means ``knowing that a stalled car can coast downhill'' and level 5 ``deciding what factors to consider in selecting stocks''.
Level 2 of ``originality'' means ``using a credit card to open a locked door'' and level 6 ``inventing a new type of man-made fiber''.
Every point represents one job. 
A higher level of either of these two abilities correlates, as expected,  with a lower probability to be automated.
% To illustrate these levels, we refer to the anchors provided by \onetonline.
% 
% 
% The correlation is -0.75 and -0.67, respectively.
% The third graph depicts the probability that a job can be automated over the level of the over the level of the work ability ``Thinking Creatively'' in this job. The correlation is again very high with -0.65.
\label{fig:scatterCorr}}
\end{figure*}

% We also look at this problem from a different perspective.
In addition to inspecting every task of every job, we consider a broader picture.
We do this by looking at general properties of a job that correlate with the probability that it can be automated.
\subsection{Tasks}
\begin{figure}[htb]
% \subfloat[Probability that a task can be automated over the share of a task.]{

\centering
\includegraphics[width=.9\linewidth]{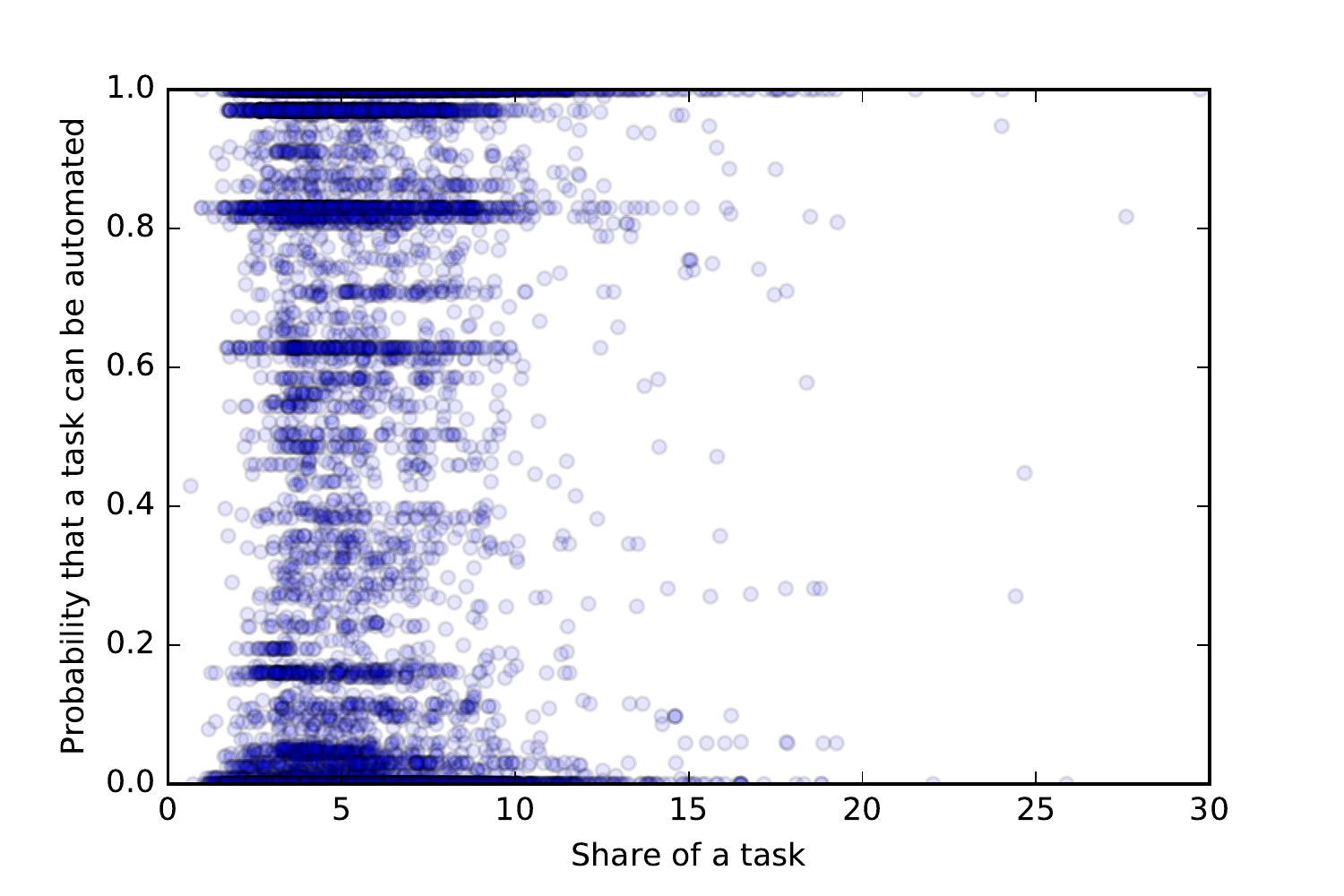}

% }
% \hfill
% \subfloat[Probability that a task can be automated over the importance of a task as defined by \onetonline.]{
% 
% \includegraphics[width=.45\linewidth]{./prob_over_IM.pdf}
% 
% }

\caption{Our results show almost no correlation between the share of a task and its probability that it will be automated.
% The correlation between the importance as stated by \onetonline{} and the probability is with $0.092$ also very small.
\label{fig:ftimoverprob}}

\end{figure}

We first analyze the share of a task.
The higher the share, the more often a task is performed.
Hence, from a machine learning perspective this means that much more training data is available.
This might lead to the conclusion that such a task is easier to automate.
To disprove this claim, we plotted the share of a task over the probability that a task can be automated according to our linear program.
The resulting graph is shown in Figure~\ref{fig:ftimoverprob}.
Every dot represents one task, with its share on the $x$-axis and its probability on the $y$-axis.
We see that there is barely any correlation between these two.
% The correlation is indeed only $0.098$.
We conclude that tasks that are done more frequently are not more likely to be automated.
% \philipp{remove the tasks here?}

% \Onetonline{} provides us with an importance of a task.
% Note that their importance is not normalized.
% We do not expect that the importance of a task positively or negatively influences the probability that it can be automated; a hypothesis that is confirmed in Figure~\ref{fig:ftimoverprob} (the correlation is only  $0.092$).

\subsection{Jobs}
We continue our analysis by looking at the correlation between the properties that a job has, e.g., what kind of degree is necessary to do a job, and the probability that this job can be automated.
Correlation does not imply causation, but nevertheless, these results reveal some interesting nuggets.

\begin{figure}[htb]
% \subfloat[Degree of Automation]{
\centering
 \includegraphics[width=.9\linewidth]{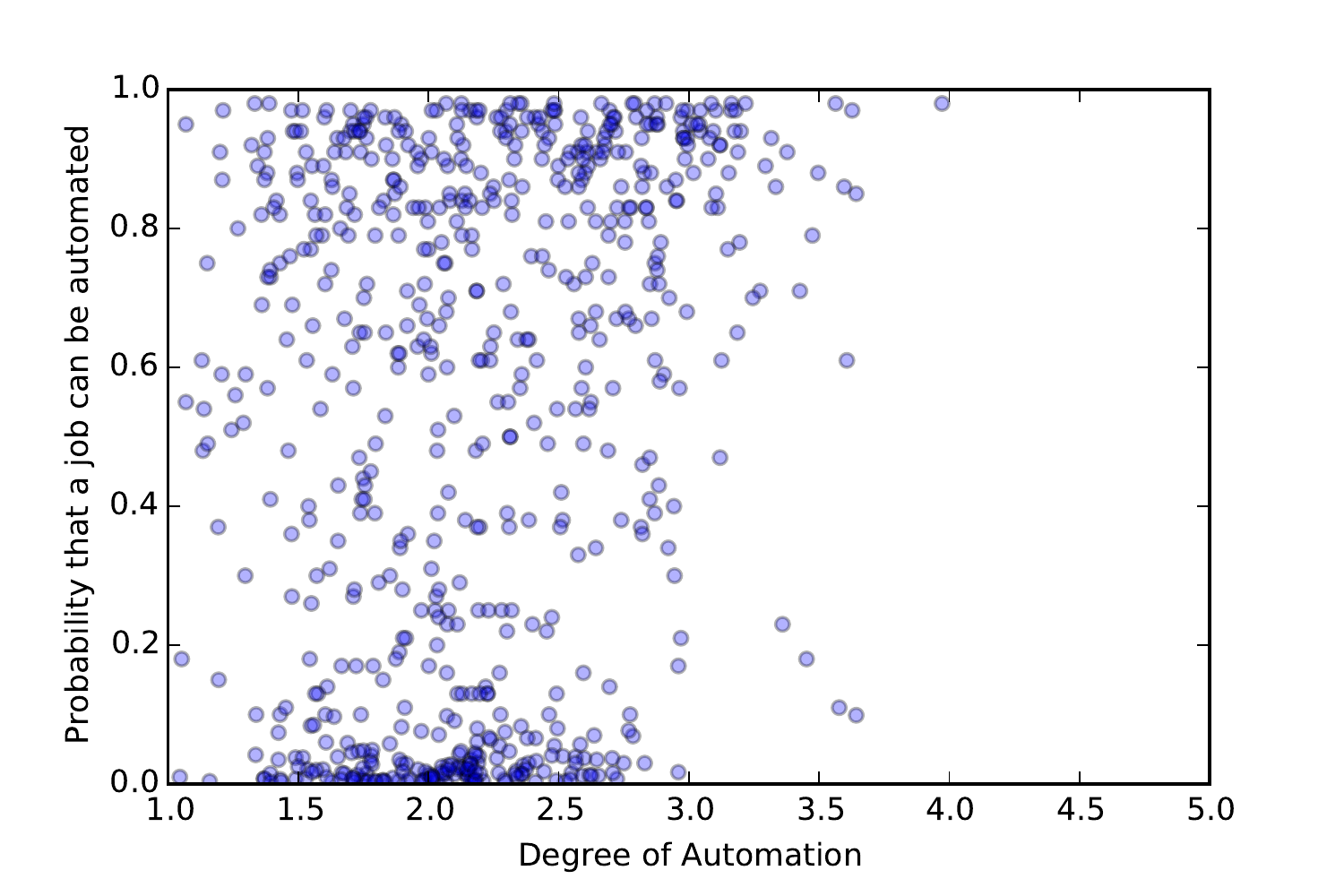}

% \hfill
% \subfloat[email]{
% 
%  \includegraphics[width=.45\linewidth]{./ElectronicMail.pdf}
% 
% }

\caption{The probability that a job can be automated over the level of the current ``Degree of Automation''.
This level ranges from $1$ (Not at all automated) to $5$ (Completely automated).
The rather small correlation of $0.23$ implies that different jobs will soon be affected.
% The probability that a job can be automated over the level of ``Electronc Mail''. 
% The level ranges from 1 (Never) to 5 (Every day). 
% The correlation is $-0.54$.
% We want to point out that correlation does not imply causation.
\label{fig:scatterDegreeEmail}
}

\end{figure}

% Let us start with a few to be expected correlations.
\Onetonline{} provides the level that the ability ``deductive reasoning'' is used in a job.
The level ranges from $1$ to $7$, where for example level 2 means ``knowing that a stalled car can coast downhill'' and level 5 ``deciding what factors to consider in selecting stocks''.
% These 
For every job, we have one value between $1$ and $7$.
% This allows us to create a scatter plot.
% We plot the probability that a job can be automated over the level of deductive reasoning needed in a job.
The resulting graph can be seen in Figure~\ref{fig:scatterCorr}.
Every job is represented by one dot; its $x$-coordinate being its level and the $y$-coordinate its probability.
We can see that jobs that require a high level of deductive reasoning tend to have a lower probability of being automated.
\begin{figure*}[htb]
\subfloat[On the Job Training]{

 \includegraphics[width=.4\linewidth]{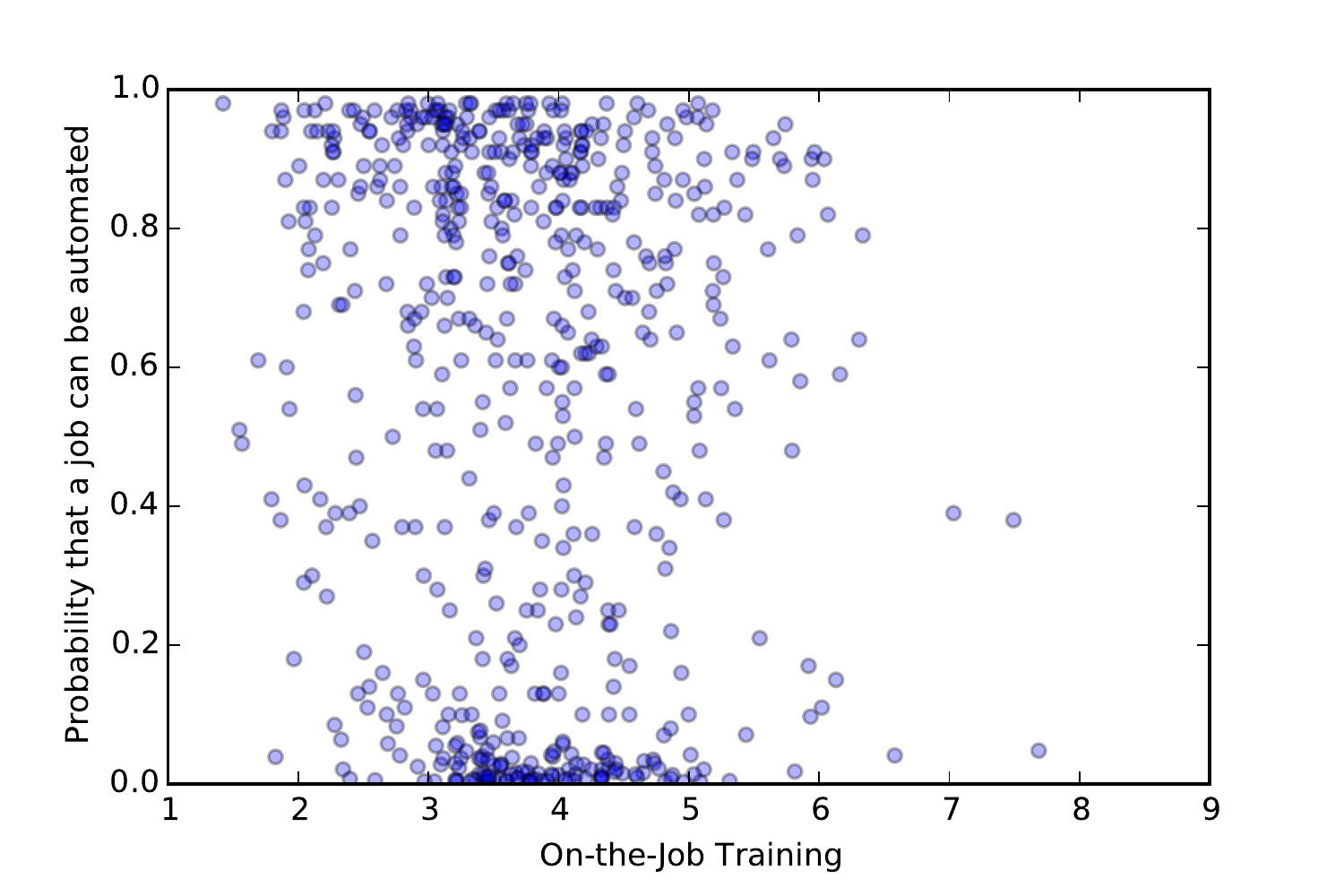}

}
\hfill
\subfloat[Required Level of Education]{

 \includegraphics[width=.4\linewidth]{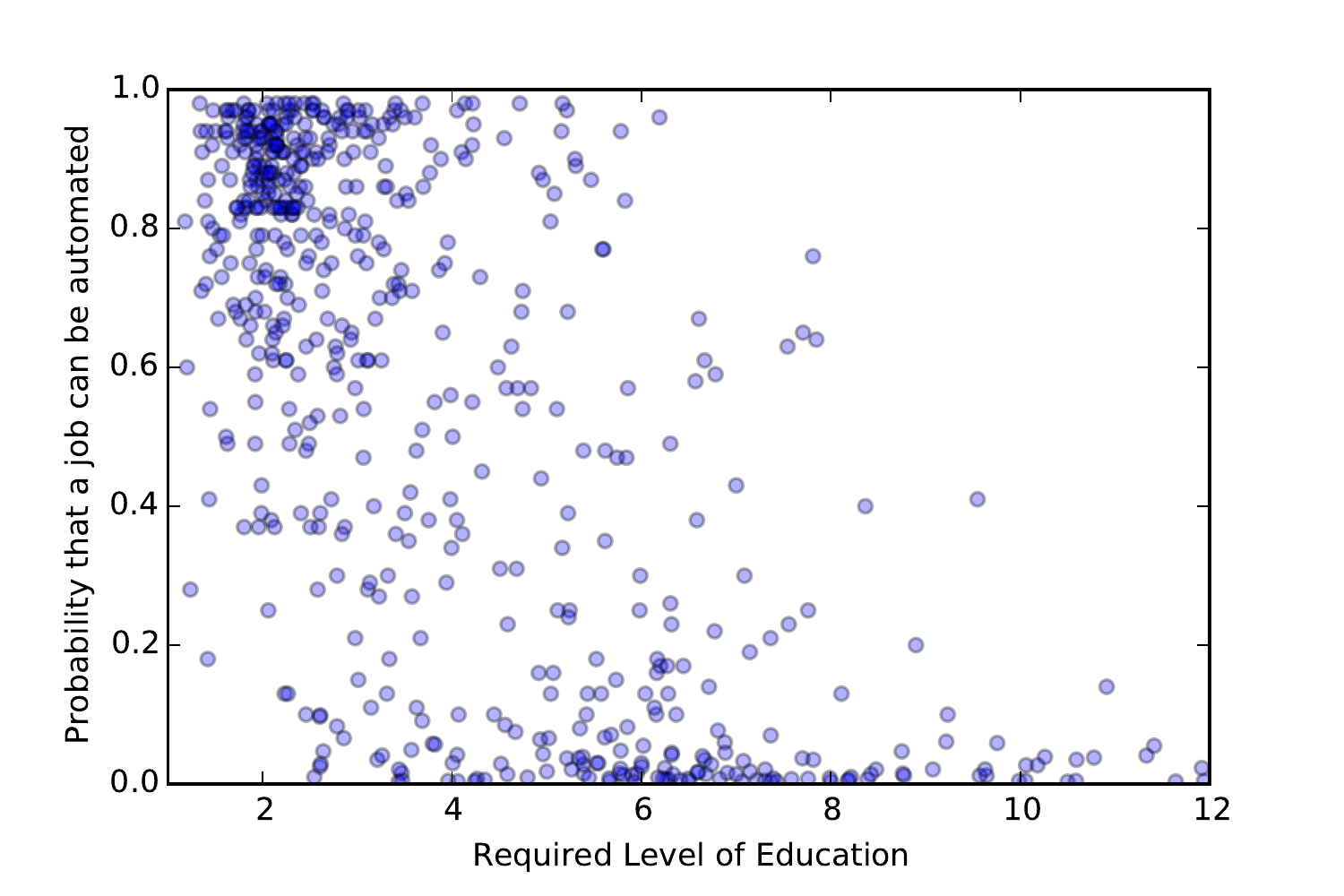}

}

\caption{
The probability that a job can be automated over the amount of ``On the Job Training'' (ranging from 1 (none or short demonstration) to 9 (over 10 years) and the ``Required of Level of Education'' (ranging from 1 (less than a high school diploma) to 12 (post-doctoral training)). 
% The correlation is $-0.092$ and $-0.68$, respectively.
\label{fig:scatterJobEducation}}

\end{figure*}
A similar result can be seen for ``originality'' (see Figure~\ref{fig:scatterCorr}).
Level 2 of ``originality'' means ``using a credit card to open a locked door'' and level 6 means ``inventing a new type of man-made fiber''.
This confirms our expectation that these abilities will remain difficult for a computer.

\Onetonline{} even has an explicit value for the current level of the ``degree of automation'' for each job.
This level ranges from $1$ (not at all automated) to $5$ (completely automated).
As depicted in Figure~\ref{fig:scatterDegreeEmail}, the already existing level of automation barely correlates with the probability that this job will be automated.
% This indicates that jobs that are already affected by this change do not have a higher probability to be automated.
This indicates that not only jobs that are already affected by automation are in danger, but also a whole new set of jobs.
This is aligned with the recent worries about many new jobs soon being affected by computerization.

% An unexpected result can be seen for how often ``electronic mail'' is used in the work context of a job.
% Figure~\ref{fig:scatterDegreeEmail} shows that there is a negative correlation between using email and the probability that a job can be automated.
% The top left corner (high probability that a job can be automated and emails are used at most once a month) and the bottom right corner (low probability that a job can be automated and emails are used daily) contain a lot of jobs.
% This should remind us that we look at correlation.
% We do not claim that using email often implies that a job is difficult to automate.

We conclude this section by looking at the effect that the level of required education for a job has on the probability to be automated.
% The level of education correlates negatively with the probability to be automated.
Jobs that require only very little education (level 1, i.e., less than a high school diploma) tend to have a higher probability than jobs that require an associate degree (level 5) which in turn have a higher probability than jobs that require post-doctoral training (level 12).
Most jobs that require little education are in danger.
It is noteworthy that the effect of training before the job is much stronger than the effect of on the job training.
Jobs that require more on the job training only have a marginally smaller probability to be automated.
Both plots are shown in Figure~\ref{fig:scatterJobEducation}.

% Last, but not least we want to look at the effect of education.
% There exists, as expected, a strong correlation between longer education and a smaller probability that a chance will be automated (see Figure~\ref{fig:scatterJobEducation}).
% Jobs with a required level of education of at least 5 (``Associate's Degree (or other 2-year degree)'') have a much lower probability to be automated.
% This effect is much stronger than the on the job training required later on (see Figure~\ref{fig:scatterJobEducation}).
% Even jobs that require ``Over 6 months, up to and including 1 year'' of on the job training (level 5), do not exhibit a similarly strong correlation. 

\section{Conclusion}

We believe that automation is one of the main challenges for society. 
In our opinion, the seminal work of Frey and Osborne did an excellent job of getting the discussion going. 
In this paper we dug a bit deeper, by looking not only at jobs -- but at the tasks that make up a job.
We hope that opening the Frey/Osborne black box will help the discussion. 
The professionals that are actually doing a job are the main experts to decide what parts of their job can or cannot be computerized. 
The Frey/Osborne work only tells these experts that their job is 87\% automatable, but what does it actually mean? 
With our work, job experts can look inside the box, and understand which tasks of their job are at risk. 
Our hope is that the job experts have a discussion which results are believable and which are not, and why.
To facilitate this discussion, we have created a web page
(\url{http://jobs-study.ethz.ch}) that allows users to comment upon our results.

\bibliographystyle{alpha} 
\newcommand{\etalchar}[1]{$^{#1}$}

\end{document}